\documentclass[prl,twocolumn,showpacs,floats]{revtex4}

\usepackage{graphicx}

\begin{document}

\title{Monotonic d-wave Superconducting Gap in Optimally-Doped Bi$_2$Sr$_{1.6}$La$_{0.4}$CuO$_6$
Superconductor by Laser-Based Angle-Resolved Photoemission Spectroscopy}
\author{Jianqiao Meng$^{1}$, Wentao Zhang$^{1}$, Guodong  Liu$^{1}$,
Lin Zhao$^{1}$, Haiyun Liu$^{1}$, Xiaowen Jia$^{1}$,Wei Lu$^{1}$,
Xiaoli Dong$^{1}$,   Guiling Wang$^{2}$, Hongbo Zhang$^{2}$, Yong
Zhou$^{2}$,  Yong  Zhu$^{3}$, Xiaoyang Wang$^{3}$, Zhongxian
Zhao$^{1}$, Zuyan Xu$^{2}$, Chuangtian Chen$^{3}$, X. J.
Zhou$^{1,*}$}

\affiliation{
\\$^{1}$National Laboratory for Superconductivity, Beijing National Laboratory for Condensed
Matter Physics, Institute of Physics, Chinese Academy of Sciences,
Beijing 100190, China
\\$^{2}$Key Laboratory for Optics, Beijing National Laboratory for Condensed Matter Physics,
Institute of Physics, Chinese Academy of Sciences, Beijing 100190,
China
\\$^{3}$Technical Institute of Physics and Chemistry, Chinese Academy of Sciences, Beijing 100190, China}
\date{April 23, 2008}
%
%

\begin{abstract}

The momentum and temperature dependence of the superconducting gap
and pseudogap in optimally-doped Bi$_2$Sr$_{1.6}$La$_{0.4}$CuO$_6$
superconductor is investigated by super-high resolution laser-based
angle-resolved photoemission spectroscopy. The measured energy gap
in the superconducting state exhibits a standard {\it d}-wave form.
Pseudogap opens above T$_c$ over a large portion of the Fermi
surface with a ``Fermi arc" formed near the nodal region. In the
region outside of the ``Fermi arc", the pseudogap has the similar
magnitude and momentum dependence as the gap in the superconducting
state which changes little with temperature and shows no abrupt
change across T$_c$. These observations indicate that the pseudogap
and superconducting gap are closely related and favor the picture
that the pseudogap is a precursor to the superconducting gap.

\end{abstract}


\maketitle

The high temperature cuprate superconductors are characterized by
their unusual superconducting state, manifested by the anisotropic
superconducting gap with predominantly {\it d}-wave
symmetry\cite{DWaveSCGap},  as well as the anomalous normal state,
exemplified by the existence of a pseudogap above the
superconducting transition temperature (T$_c$)\cite{Pseudogap}. The
origin of the pseudogap and its relation with the superconducting
gap are critical issues in understanding the mechanism of
superconductivity and exotic normal state
properties\cite{PseudogapTheory,Millis}. It has been a long-standing
debate on whether the pseudogap is intimately related to the
superconducting gap like a precursor of
pairing\cite{ChRenner,NormanArc,WangNernst,KanigelPRL} or it
originates from other competing orders that has no direct bearing on
superconductivity\cite{Deutscher,Sacuto,KTanaka,Hudson}.

Angle-resolved photoemission spectroscopy (ARPES), as a powerful
tool to directly measure the magnitude of the energy gap, has
provided key insights on the superconducting gap and pseudogap in
cuprate superconductors\cite{TwoReviews}. Recently, great effort has
been focused on investigating their relationship but the results are
split in supporting two different
pictures\cite{KTanaka,TKondo,WSLee,KanigelPRL,Terashima2C,Shi1C}. In
one class of ARPES experiments, distinct doping and temperature
dependence of the energy gap between the nodal and antinodal regions
are reported\cite{KTanaka,WSLee} which are used to support ``two
gap" picture where the pseudogap and the superconducting gap are
loosely related or independent.  Additional support comes from the
unusual gap form measured in the superconducting
state\cite{TKondo,Terashima2C}. Its strong deviation from the
standard {\it d}-wave form is interpreted as composing of ``two
components": a ``true" d-wave superconducting gap and the remanent
pseudogap that is already present in the normal
state\cite{TKondo,Terashima2C}. In another class of experiments that
supports ``one-gap" picture where the pseudogap is a precursor of
the superconducting gap, the gap in the superconducting state is
found to be consistent with a standard {\it d}-wave
form\cite{KanigelPRL,Shi1C}. Slight deviation in the underdoped
regime is interpreted as due to high-harmonic pairing
terms\cite{MesotPRL}.

\begin{figure*}[floatfix]
\begin{center}
\includegraphics[width=1.82\columnwidth,angle=0]{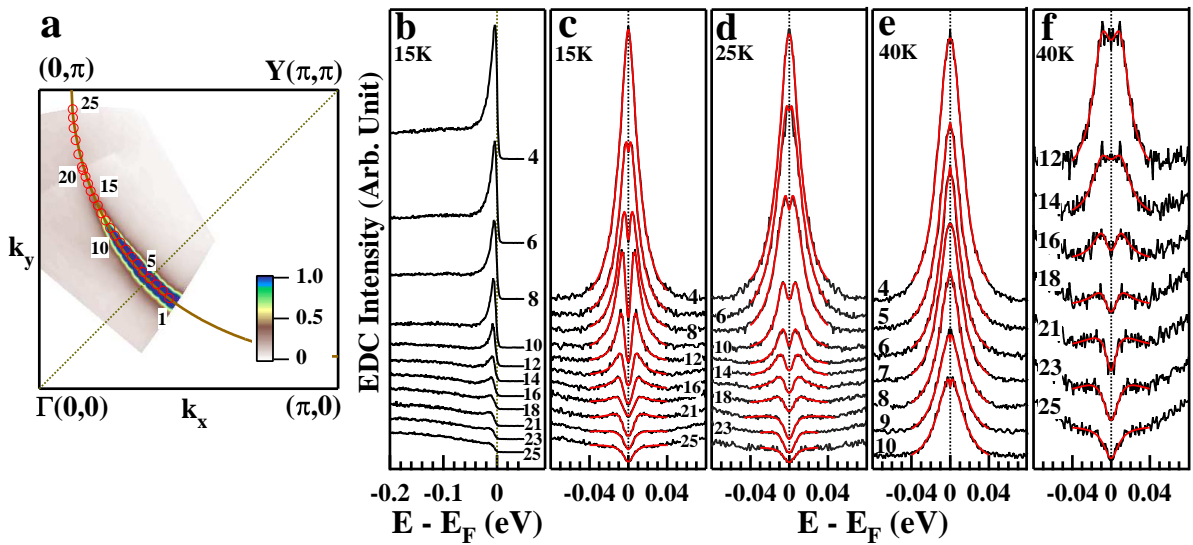}
\end{center}
\caption{Fermi surface of the optimally-doped La-Bi2201 (T$_c$=32 K)
and corresponding photoemission spectra (EDCs) on the Fermi surface
at various temperatures. (a).Spectral weight as a function of
two-dimensional momentum (k$_x$, k$_y$) integrated over [-5meV,5meV]
energy window with respect to the Fermi level E$_F$. The measured
Fermi momenta are marked by red empty circles and labeled by
numbers; (b). Original EDCs along the Fermi surface measured at 15
K. The symmetrized EDCs along the Fermi surface are shown in (c) for
15 K, (d) for 25K  and (e and f) for 40 K. The numbers on panels
(b-f) corresponds to the Fermi momentum numbers in (a). }
\end{figure*}

In light of the controversy surrounding the relationship between the
pseudogap and superconducting gap and its importance in
understanding high-T$_c$ superconductivity, we report in this paper
detailed momentum and temperature dependence of the superconducting
gap and pseudogap in Bi$_2$Sr$_{1.6}$La$_{0.4}$CuO$_6$ (La-Bi2201)
superconductor by super-high resolution laser-based ARPES
measurements.  In the superconducting state we have identified an
anisotropic energy gap that is consistent with a standard {\it
d}-wave form. This is significantly different from the previous
results on a similar superconductor\cite{TKondo}. In the normal
state, we have observed pseudogap opening with a small ``Fermi arc"
formed near the nodal region. Outside of the "Fermi arc", the
pseudogap in the normal state has the similar magnitude and momentum
dependence as the gap in the superconducting state: detailed
temperature dependence shows that the pseudogap evolves smoothly
into the superconducting gap with no abrupt change across T$_c$.
These results point to an intimate relationship between the
pseudogap and the superconducting gap which is in favor of the
``one-gap" picture that pseudogap is a precursor to the
superconducting gap.

The ARPES measurements are carried out on our newly-developed Vacuum
Ultraviolet(VUV) laser-based angle-resolved photoemission system
with advantages of super-high energy resolution, high momentum
resolution, high photon flux and enhanced bulk
sensitivity\cite{GDLiu}. The photon energy is 6.994 eV with a
bandwidth of 0.26 meV and the energy resolution of the electron
energy analyzer (Scienta R4000) was set at 0.5 meV, giving rise to
an overall energy resolution of 0.56 meV. The angular resolution is
$\sim$0.3$^{\circ}$, corresponding to a momentum resolution
$\sim$0.004${\AA}$$^{-1}$ at the photon energy of 6.994 eV. The
optimally doped Bi$_2$Sr$_{2-x}$La$_{x}$CuO$_{6}$ (La-Bi2201)(x=0.4,
T$_c$$\sim$32 K, transition width $\sim$2 K) single crystals were
grown by the traveling solvent floating zone method\cite{JianqiaoM}.
One advantage of choosing La-Bi2201 system lies in its relatively
low superconducting transition temperature that is desirable in
investigating the normal state behavior with suppressed thermal
broadening of photoemission spectra. The samples are cleaved
\emph{in situ} in vacuum with a base pressure better than
4$\times$10$^{-11}$ Torr.

Fig. 1(a) shows the Fermi surface mapping of the optimally doped
La-Bi2201 (T$_c$=32 K) measured at 15 K. The low photon energy and
high photon flux have made it possible to take dense sampling of the
measurements in the momentum space. The photoemission spectra
(Energy Distribution Curves, EDCs) along the Fermi surface are
plotted in Fig. 1(b).  The EDCs near the nodal region show sharp
peaks that are similar to those observed in Bi2212\cite{ZhangPRL}.
When the momentum moves away from the nodal region to the (0,$\pi$)
antinodal region, the EDC peaks get weaker, but peak feature remains
along the entire Fermi surface even for the one close to the
antinodal region. The EDC peak position also shifts away from the
Fermi level when the momentum moves from the nodal to the antinodal
region, indicating a gap opening in the superconducting state.  Note
that the EDCs near the antinodal region do not show any feature near
40 meV that was reported in a previous measurement\cite{TKondo}.

In order to extract the energy gap, we have symmetrized the original
EDCs with respect to the Fermi level, as shown in Fig. 1c for the 15
K measurements, and Fig. 1d and Fig. 1(e-f) for 25 K and 40 K,
respectively. The symmetrization procedure not only provides an
intuitive way in visualizing the energy gap, but also removes the
effect of Fermi cutoff in photoemission spectra and provides a
quantitative way in extracting the gap size\cite{MRNorman}. The
symmetrized EDCs have been fitted using the general phenomenological
form\cite{MRNorman}; the fitted curves are overlaid in Fig. 1(c-f)
and the extracted gap size is plotted in Fig. 2.

\begin{figure}[tbp]
\begin{center}
\includegraphics[width=0.9\columnwidth,angle=0]{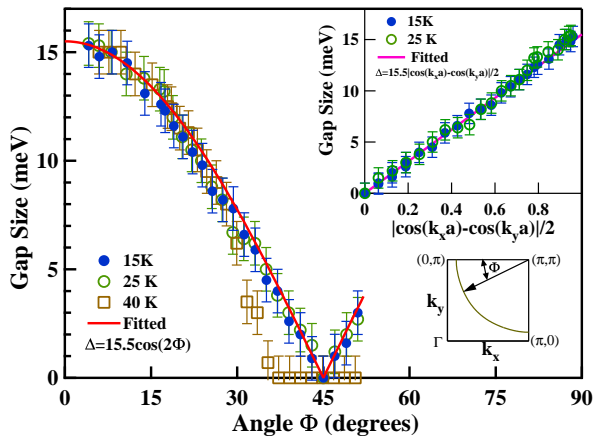}
\end{center}
\caption{Energy gap along the Fermi surface measured at 15 K (solid
circles), 25 K (empty circles) and 40 K (empty squares) on the
optimally-doped La-Bi2201 (T$_c$=32 K). The solid red line is fitted
from the measured data at 15 K which gives
$\Delta$=15.5cos(2$\Phi$). The $\Phi$ angle is defined as shown in
the bottom-right inset.  The upper-right inset shows the gap size as
a function of $|cos(k_xa)-cos(k_ya)|/2$ at 15 K and 25 K. The pink
line represents a fitted line with
$\Delta$=15.5$|cos(k_xa)-cos(k_ya)|/2$.
 }
\end{figure}

As shown in Fig. 2, the gap in the superconducting state exhibits a
clear anisotropic behavior that is consistent with a standard {\it
d}-wave form $\Delta$=$\Delta$$_0$cos(2$\Phi$) (or in a more strict
sense, $\Delta$=$\Delta$$_0$$|cos(k_xa)-cos(k_ya)|$/2 form as shown
in the inset of Fig. 2) with a maximum energy gap $\Delta$$_0=$15.5
meV. It is also interesting to note that the gap is nearly identical
for the 15 K and 25 K measurements for such a T$_c$=32 K
superconductor. These results are significantly different from a
recent measurement where the gap in the superconducting state
deviates strongly from the standard {\it d}-wave form with an
antinodal gap at 40 meV\cite{TKondo}.  An earlier
measurement\cite{JHarris} gave an antinodal gap at 10$\sim$12 meV
which is close to our present measurement, but it also reported
strong deviation from the standard {\it d}-wave form. While the
non-{\it d}-wave energy gap can be interpreted as composed of two
components in the previous measurement\cite{TKondo}, our present
results clearly indicate that the gap in the superconducting state
is dominated by a {\it d}-wave component.

In the normal state above T$_c$=32 K, the Fermi surface measured at
40 K is still gapped over a large portion except for the section
near the nodal region that shows a zero gap, as seen from the
symmetrized EDCs (Fig. 1e-f for 40 K) and the extracted pseudogap
(40 K data in Fig. 2). This is consistent with the ``Fermi arc"
picture observed in other high temperature
superconductors\cite{NormanArc,KTanaka,CampuzanoNP}. Note that the
pseudogap outside of the ``Fermi arc" region shows similar magnitude
and momentum dependence as the gap in the superconducting state
(Fig. 2).

\begin{figure}[tbp]
\begin{center}
\includegraphics[width=0.9\columnwidth,angle=0]{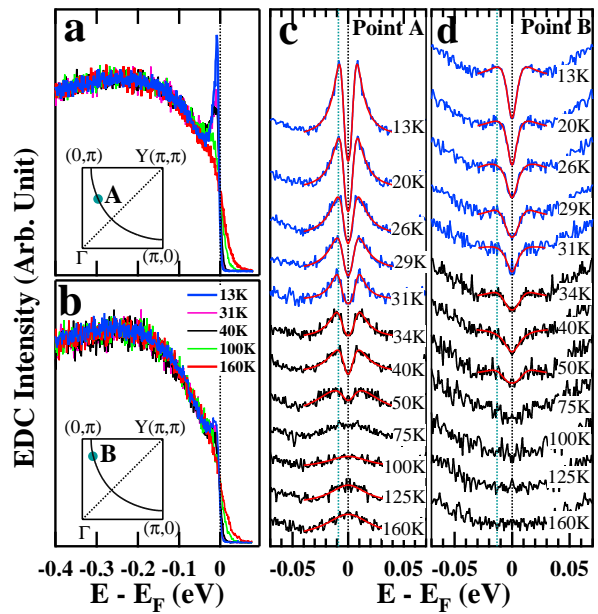}
\end{center}
\caption{(a,b). Temperature dependence of representative EDCs at two
Fermi momenta on the Fermi surface in optimally-doped La-Bi2201. The
location of the Fermi momenta is indicated in the inset. Detailed
temperature dependence of the symmetrized EDCs for the Fermi
momentum A are shown in (c) and for the Fermi momentum B in (d). The
dashed lines in (c) and (d) serve as a guide to the eye. }
\end{figure}

Fig. 3 shows detailed temperature dependence of EDCs and the
associated energy gap for two representative momenta on the Fermi
surface. Strong temperature dependence of the EDCs is observed for
the Fermi momentum A (Fig. 3a). At high temperatures like 100 K or
above, the EDCs show a broad hump structure near -0.2 eV with no
observable peak near the Fermi level. Upon cooling, the high-energy
-0.2 eV broad hump shows little change with temperature, while a new
structure emerges near the Fermi level and develops into a sharp
``quasiparticle" peak in the superconducting state, giving rise to a
peak-dip-hump structure in EDCs. This temperature evolution and
peak-dip-hump structure are reminiscent to that observed in other
high temperature superconductors like Bi2212\cite{FedorovPRL}. When
moving towards the antinodal region, as for the Fermi momentum B
(Fig. 3b),  the EDCs qualitatively show similar behavior although
the temperature effect gets much weaker. One can still see a weak
peak developed at low temperatures, e.g., 13 K,  near the Fermi
level.

\begin{figure}[tbp]
\begin{center}
\includegraphics[width=0.9\columnwidth,angle=0]{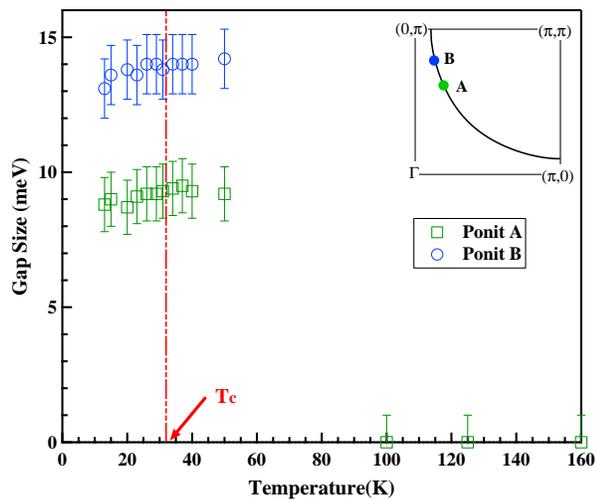}
\end{center}
\caption{Temperature dependence of the energy gap for two Fermi
momenta A (empty squares) and B (empty circles) as indicated in
insets of Fig. 3(a) and (b), and also indicated in the up-right
inset, for optimally-doped La-Bi2201. The dashed line indicates
T$_c$=32 K.
 }
\end{figure}

To examine the evolution of the energy gap with temperature, Fig. 3c
and 3d show symmetrized EDCs measured at different temperatures for
the Fermi momenta A and B, respectively. The gap size extracted by
fitting the symmetrized EDCs with the general formula\cite{MRNorman}
are plotted in Fig. 4. For the Fermi momentum A, as seen from Fig.
3c, signature of gap opening in the superconducting state persists
above T$_c$=32 K, remaining obvious at 50 K, getting less clear at
75 K, and appear to disappear around 100 K and above as evidenced by
the appearance of a broad peak. The gap size below 50 K (Fig. 4)
shows little change with temperature and no abrupt change is
observed across T$_c$. The data at 75 K is hard to fit to get a
reliable gap size, thus not included in Fig. 4. When the momentum
moves closer to the antinodal region, as for the Fermi momentum B,
similar behaviors are observed, i.e., below 50 K, the gap size is
nearly a constant without an abrupt change near T$_c$.  But in this
case, different from the Fermi momentum A, there is no broad peak
recovered above 100 K, probably indicating a higher pseudogap
temperature. This is qualitatively consistent with the
transport\cite{AndoRMapping} and NMR\cite{ZhengNMR} measurements on
the same material that give a pseudogap temperature between
100$\sim$150 K.

From precise gap measurement, there are clear signatures that can
distinct between ``one-gap" and ``two-gap" scenarios\cite{Millis} .
In the ``two-gap" picture where the pseudogap and superconducting
gap are assumed independent, because the superconducting gap opens
below T$_c$ {\it in addition to} the pseudogap that already opens in
the normal state and persists into the superconducting state, one
would expect to observe two effects: (1). Deviation of the energy
gap from a standard {\it d}-wave form in the superconducting state
with a possible break in the measured gap form\cite{TKondo}; (2).
Outside of the ``Fermi arc" region, one should expect to see an
increase in gap size in the superconducting state.  Our observations
of standard {\it d}-wave form in the superconducting state (Fig. 2),
similar magnitude and momentum dependence of the pseudogap and the
gap in the superconducting state outside of the ``Fermi arc" region
(Fig. 2),  smooth evolution of the gap size across T$_c$ and no
indication of gap size increase upon entering the superconducting
state (Fig. 4), are not compatible with the expectations of the
``two-gap" picture.  They favor the ``one-gap" picture where the
pseudogap and superconducting gap are closely related and the
pseudogap transforms into the superconducting gap across T$_c$. Note
that, although  the region outside of the ``Fermi arc" shows little
change of the gap size with temperature (Fig. 4), the EDCs exhibit
strong temperature dependence with a ``quasiparticle" peak developed
in the superconducting state(Fig. 3a and 3b) that can be related
with the establishment of phase
coherence\cite{FedorovPRL,KanigelPRL}. This suggests that the
pseudogap region on the Fermi surface can sense the occurrence of
superconductivity through acquiring phase coherence.

In conclusion, from our precise measurements on the detailed
momentum and temperature dependence of the energy gap in optimally
doped La-Bi2201, we provide clear evidence to show that the
pseudogap and superconducting gap are intimately related. Our
observations are in favor of the ``one-gap" picture that the
pseudogap is a precursor to the superconducting gap and
superconductivity is realized by establishing a phase coherence.

We acknowledge helpful discussions with T. Xiang. This work is
supported by the NSFC(10525417 and 10734120), the MOST of China (973
project No: 2006CB601002, 2006CB921302), and CAS (Projects ITSNEM
and 100-Talent).

$^{*}$Corresponding author: XJZhou@aphy.iphy.ac.cn

\newpage

\end{document}